\documentclass[aps,prl,showpacs,twocolumn,superscriptaddress,floatfix]{revtex4}
\usepackage{graphicx}
\usepackage{amssymb,amsmath} 
\usepackage{color}
\usepackage{multirow}
\usepackage[normalem]{ulem}

\begin{document}
\preprint{}

\title{Giant magnetization canting due to symmetry breaking in zigzag Co chains on Ir(001)}

\newcommand{\hamburg}{Institute of Applied Physics, Hamburg University, Jungiusstrasse 11, D-20355 Hamburg, Germany}
\newcommand{\kiel}{Institute of Theoretical Physics and Astrophysics, Christian-Albrechts University of Kiel, Leibnizstrasse 15, D-24098 Kiel, Germany}
\newcommand{\yuriy}{Peter Gr\"unberg Institut and Institute for Advanced Simulation, Forschungszentrum J\"ulich and JARA,
D-52425 J\"ulich, Germany}
\newcommand{\mhc}{Current address: Department of Physics, Mount Holyoke College, South Hadley, MA 01075, USA}

\author{Bertrand Dup\'e}
\email{dupe@theo-physik.uni-kiel.de}
\affiliation{\kiel}
\author{Jessica E. Bickel}
\email{jbickel@mtholyoke.edu}
\altaffiliation{\mhc}
\affiliation{\hamburg}
\author{Yuriy Mokrousov}
\affiliation{\yuriy}
\author{Fabian Otte}
\affiliation{\kiel}
\author{Kirsten von Bergmann}
\affiliation{\hamburg}
\author{Andr\'e Kubetzka}
\affiliation{\hamburg}
\author{Stefan Heinze}
\affiliation{\kiel}
\author{Roland Wiesendanger}
\affiliation{\hamburg}

\date{\today}

\begin{abstract}
We demonstrate a canted magnetization of biatomic zigzag Co chains grown on the ($5 \times 1$) reconstructed
Ir(001) surface using density functional theory calculations and spin-polarized scanning tunneling microscopy (SP-STM) experiments.
Biatomic Co chains grow in three different structural configurations and are all in a ferromagnetic state. Two chain types possess
high symmetry due to two equivalent atomic strands and an easy magnetization direction which is along one of the
principal crystallographic axes. The easy magnetization axis of the zigzag Co chains is canted away from the surface
normal by an angle of $33^\circ$. This giant effect is caused by the broken chain symmetry on the substrate
in combination with the strong spin-orbit coupling of Ir.
SP-STM measurements confirm stable ferromagnetic order of the zigzag chains with a canted magnetization.
\end{abstract}


\maketitle

Low dimensional magnetic nanostructures at surfaces such as single atoms, clusters, and atomic chains
constitute model systems to explore spintronic concepts
at the ultimate scale. In view of their enhanced magnetocrystalline anisotropy energy (MAE) such nanomagnets are attractive
for further miniaturization of data storage as the MAE acts as a barrier and stabilizes the magnetization against thermal
fluctuations.
One of the most striking examples is the giant anisotropy reported for single atom Co chains grown at the step edges of a Pt(111)
surface allowing the observation of ferromagnetic order \cite{gambardella2002}. Due to the step edge
the easy magnetization axis is not oriented along one of the high symmetry axis but it is canted from the surface normal
by about 43$^{\circ}$ towards the upper Pt terrace \cite{gambardella2002} which is caused by the competition of contributions
to the magnetocrystalline anisotropy from the Co and Pt atoms \cite{Ujf2004,PhysRevB.69.212410,Baud_PRB_2006}.
Addition of more atomic Co strands at the step edge reorients the magnetization direction away from the upper terrace
and the direction oscillates until it is nearly perpendicular to the vicinal surface for a coverage of a monolayer
\cite{gambardella2004,baud2006ss}.
Since these pioneering experiments were performed, quasi one-dimensional chains at surfaces have been the subject of intense research,
in particular in theoretical
studies
(e.g.~\cite{Ujf2004,PhysRevB.69.212410,Baud_PRB_2006,baud2006ss,SpisakSS2003,SpisakPRB2003,Mok2007,Lounis2008,MazPRB2009,MokrousovPRB2009,Has2010}).
However, very few other systems have been
characterized experimentally concerning their magnetic state~\cite{hirjibehedin2006spi,mmmPrl2012,Loth13}.

\begin{figure}[htb]
\includegraphics[width=8.5cm]{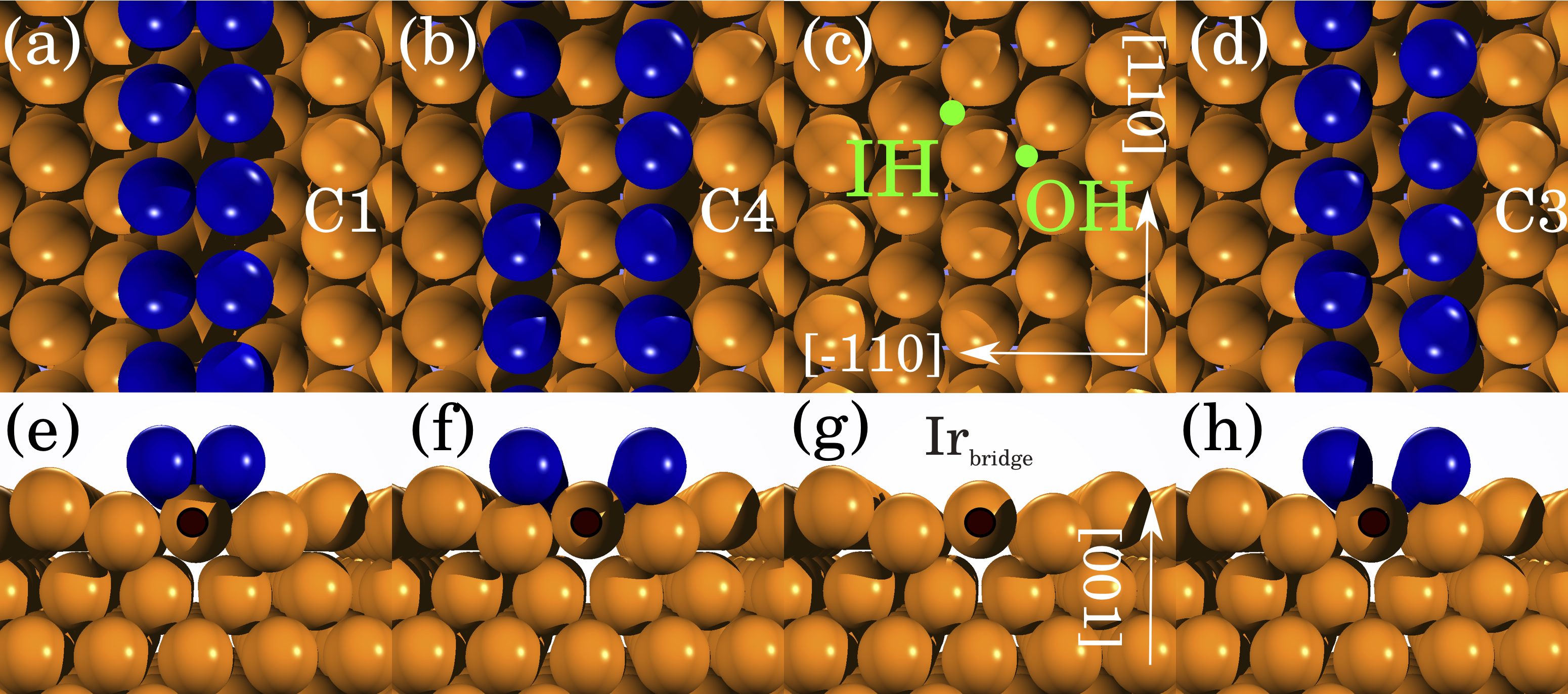}
\caption{(color online) (a-d) Top and (e-h) side view of the three different chain configurations and the (5 $\times$ 1) reconstructed Ir(001)
surface. In (c) the inner hollow (IH) and the outer hollow (OH) sites and in (g) the bridge Ir atom are marked. For the C1 (a,e)
and C4 (b,f) chains
the two strands are in the IH and OH sites, respectively, while for the C3 chain (d,h) one is in the IH and the other
in the OH site.
}
\label{fig:structure}
\end{figure}

One promising surface on which to grow quasi-one-dimensional chains is the (5 $\times$ 1) surface reconstruction of Ir(001) which
exhibits a trench structure that allows self-assembly
of different types of biatomic chains \cite{Gilarowski2000290,PhysRevB.67.12542}.
Two of the chain configurations possess a high symmetry with two equivalent atomic strands that
adsorb either on the inner or on the outer hollow site of the trench (see Figs.~\ref{fig:structure}(a,b)). For such biatomic
Fe chains it has been reported previously that due to the hybridization between Fe and Ir
the magnetic coupling and the easy magnetization axis depend significantly on the adsorption
site of the atoms \cite{MazPRB2009,MokrousovPRB2009,mmmPrl2012}. If one of the two strands of the biatomic chain adsorbs on
the inner hollow
site while the other adsorbs on the outer hollow site a zigzag chain forms which lacks a mirror plane along its axis
(see Figs.~\ref{fig:structure}(c,d)).
This type of chain has not been observed for Fe, however, as we show here it can be formed by depositing Co.

In this paper, we demonstrate that the symmetry breaking of zigzag chains on a surface can lead to a giant canting
of their easy magnetization direction. This is remarkable, since both the chain and the substrate separately possess
high symmetry and exhibit a relatively small buckling when brought in contact with each other. 
It is in clear contrast to the case of atomic Co chains at a Pt(111) step edge where the surface
structure already has broken symmetry and a canting of the magnetization is expected.

Our first-principles calculations based on density functional theory (DFT)
show that biatomic zigzag Co chains on the $(5\times1)$ reconstructed Ir(001) surface are ferromagnetic and
that their easy axis is canted from the surface normal by an angle of $33^\circ$.
We explain that this very large effect is due to the local symmetry breaking of the bridge chain of Ir atoms
in between the Co strands,
which provides the dominant contribution to the MAE and favors large canting of the magnetization. Experiments performed using spin-polarized scanning tunneling
microscopy (SP-STM) confirm ferromagnetic order at 8 K with a canted magnetization direction of the Co
zigzag chains.

We study the magnetic properties of biatomic Co chains on the (5 $\times$ 1) reconstructed Ir(001) surface
by first-principles calculations based on DFT applying
the film version of the full potential linearized augmented plane wave (FLAPW) method as implemented in the FLEUR code \cite{FLEUR}.
We used a symmetric slab consisting of 37 Ir atoms (7 substrate layers) and 4 Co atoms.
The resulting ($5 \times 1$) supercell has inversion symmetry with Co chains on both surfaces and adjacent Co chains
are separated by $13.51$ \AA.
The structural relaxation was carried out using a mixed LDA/GGA functional introduced by De Santis \emph{et al} \cite{PhysRevB.75.205432}
to treat systems of 3$d$- and 5$d$-transition metals. Further computational details can be found in \cite{suppl}.

We denote the biatomic chains with both atomic strands adsorbed in the
inner (IH) or outer hollow (OH) sites as C1 and C4, respectively \cite{SpisakSS2003},
while the zigzag chain is referred to as C3 chain (cf.~Fig.~\ref{fig:structure}).
The separation between the two Co strands increases as we move from the C1 to the C3 and to the C4 chain from 2.26~\AA\ to
3.18~{\AA} up to 4.10~{\AA}. For the zigzag (C3) chain, there is a small buckling between the two inequivalent Co strands
amounting to only 0.14~{\AA}. The magnetic moments of the Co atoms depend both on the hybridization
between the two strands and of the strands with the Ir substrate.
The resulting values for the Co atoms in C1 are $1.9~\mu_B$ and decrease to $1.8~\mu_B$ for the Co in the
C4 configuration. For the C3 chains the moments are comparable with the high symmetry configurations:
$1.9~\mu_B$ for the IH site atom and $1.8~\mu_B$ for the OH site atom. For all chain configurations
strong ferromagnetic exchange interaction is found along the chain. Due to the larger spacing between
the two strands, the ferromagnetic coupling between the two strands is reduced by one order of magnitude
for the C3 and C4 chain (values can be found in \cite{suppl}).

We calculated the MAE of the Co chains using the magnetic force theorem \cite{doi-10.1139/p80-159}
and checked several configurations using self-consistent calculations \cite{suppl}.
For the biatomic chains possessing mirror symmetries along and perpendicular to their axis, the easy axis of the magnetization
can only lie along one of the high symmetry directions: out-of-plane (OP) to the surface,
in-plane along the chain axis ($[110]$ direction) and in-plane perpendicular to the chain axis ($[\bar{1}10]$ direction)
(see Fig.~\ref{fig:structure}).
The easy magnetization axis of the C1 chain, where the two strands are closest, is out-of-plane
with energy differences of 0.17~meV/Co-atom and 0.40~meV/Co-atom with respect to the $[110]$ direction and $[\bar{1}10]$ direction,
respectively. For the C4 chain, the easy axis is oriented along the chain axis and the hard axis is along the surface normal
(1.00~meV/Co-atom) and the $[\bar{1}10]$ direction is the intermediate state (0.59~meV/Co-atom).
Qualitatively, the same results were found for biatomic Fe chains \cite{MokrousovPRB2009}.

\begin{figure}[htb]
\includegraphics[width=8.cm]{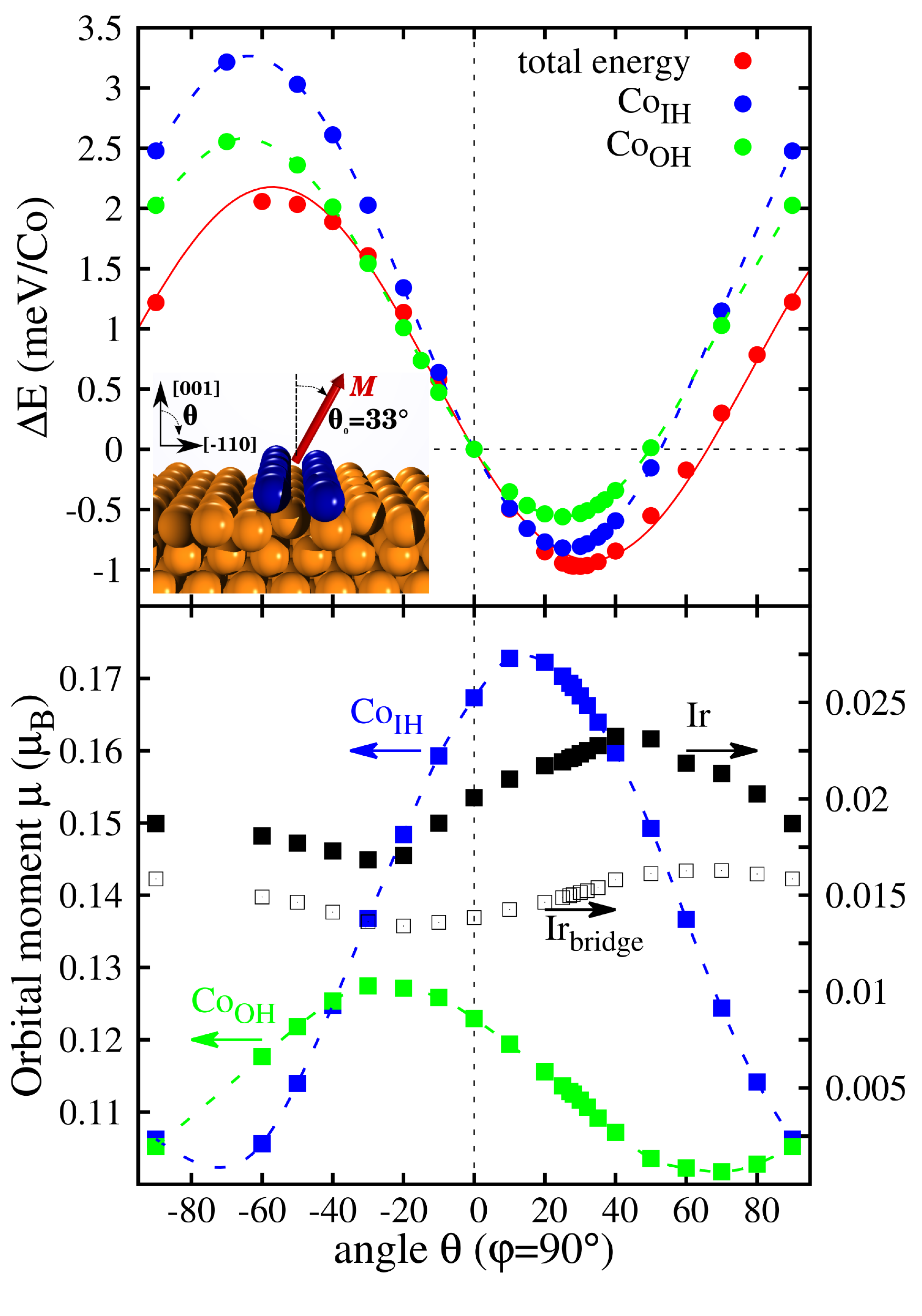} \\
\caption{(color online) (a) Magnetocrystalline anisotropy energy (MAE) calculated for the zigzag (C3) Co chain.
The SQA is rotated in a plane perpendicular to the chain axis as shown in the inset. Beside the
total energy, calculations are shown in which SOC has been turned off in one of the two Co strands
(${\rm Co}_{\rm IH}$: SOC off in the outer hollow strand and vice versa for ${\rm Co}_{\rm OH}$).
The dotted lines represent a fit to the MAE according to uniaxial anisotropy expression.
(b) Orbital moments of the two inequivalent Co atoms, all Ir atoms, and the bridge Ir atom.}
\label{fig:MAE}
\end{figure}

Due to symmetry breaking, the situation becomes more complicated for the deposited zigzag (C3)
chains and the easy magnetization direction does not need to align with a high symmetry direction, although it has to do so for the free-standing zigzag chain.
The total energy was therefore calculated by rotating the spin quantization axis (SQA) also in a plane perpendicular
to the chain axis. As can be seen in Fig.~\ref{fig:MAE}(a), a minimum of about 1 meV/Co-atom
is obtained for an angle of $\theta_0=33^{\circ}$ (see red curve).

In the spirit of the Bruno formula which links the energy minimum to the maximum of the orbital moment~\cite{PhysRevB.39.86}, we
can interpret this result based on the orbital contributions of individual atoms displayed in Fig.~\ref{fig:MAE}(b).
Since the zigzag chain is composed of two non-equivalent strands, we observe different sizes and angular dependencies
for the inner and outer hollow Co atoms. While the maximum is at a positive angle of $15^\circ$ for the inner hollow Co strand
it is at $-23^\circ$ for the outer hollow strand. This leads to an opposite preference of the favorable magnetization
direction for the two Co atoms. In agreement, the calculated energy displays a minimum of $\theta \approx 0^\circ$
if we turn off the SOC contribution from the entire substrate (not shown)~\cite{note_noco}.

The driving force behind the giant canting stems from interplay of the broken chain symmetry and the large Ir substrate contribution
to the anisotropy of total energy and orbital moments. When
SOC is switched on in the substrate and in only one of the Co strands, we still acquire a total energy minimum
at positives angles (Fig.~\ref{fig:MAE}(a)) irrespective of which Co chain is "switched on". This shows that although one of the Co strands favors canting with a negative angle, its contribution is overwhelmed by that of the substrate.
In agreement with the total energy,  the maximum orbital moment of all Ir
atoms in the first surface layer is found at a large positive angle of $44^\circ$.
The main contribution to this orbital moment comes from the chain of Ir atoms at the
bridge site between the two  Co strands (cf.~Fig.~\ref{fig:structure}). For each of the bridge Ir atoms the nearest
neighbor Co atoms, which are different in their electronic structure due to different coordination, form a triangle with identical orientation. This causes fundamental breaking of
local symmetry which results in all Ir bridge atoms favoring strong canting of magnetization
towards a giant positive angle.

\begin{figure}[htb]
\includegraphics[width=8.cm]{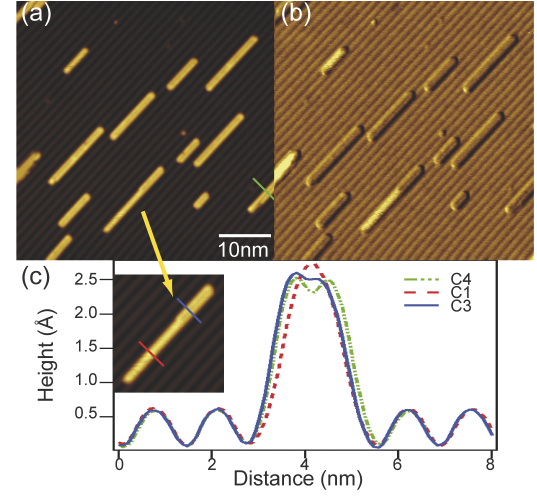}
\caption{(color online) (a) STM topography image and (b) simultaneously obtained d$I$/d$U$ image of Co chains on the
$(5 \times 1)$ Ir(001) surface.
(c) Line profiles of a C4 chain in image (a) and two profiles along a chain which changes structure along its axis
from C1 to C3 (magnified in inset).
Measurement parameters: W-tip, \textit{T}~=~7.8~K, $U_{bias}$~=~$-$60~mV, $I$~=~0.2~nA.}
\label{fig:topo}
\end{figure}

To confirm the predicted ferromagnetic order and the canted magnetization of the zigzag Co chains
we have performed spin-polarized STM experiments~\cite{wiesendanger_RevModPhys_2009}. Bulk Cr tips are used
with an arbitrary magnetization direction that is not changed by the
application of an external magnetic field. Details on the sample and tip
preparation can be found in \cite{suppl}.

Figure \ref{fig:topo}(a),(b) show STM topography and d$I$/d$U$ images of Co deposited on the (5 $\times$ 1)
reconstructed Ir(001) surface. As expected wires form along the trench structure of the surface.
Several contrast levels are present in the d$I$/d$U$ image, indicating differences in the electronic structure, which demonstrates that
multiple chain structures are present on the surface~\cite{note_defect}.
The chain configurations can be
determined from line profiles shown in Fig.~\ref{fig:topo}(c) taken perpendicular to the chains as marked in the inset and Fig.~\ref{fig:topo}(a).
Three different profiles
can be distinguished: the dashed red curve shows a single peak with a narrow chain profile, the dot-dashed green curve shows a double peak with a wide chain profile, and the solid blue line a double peak with a profile that matches the narrow chain on
the right and the wider chain on the left.
When these three profiles are compared to the possible atomic arrangements in the Ir(001)-($5 \times 1$) trench,
the chains can be identified as C1, C4, and C3, respectively.
The line profiles obtained from our DFT calculations based on the Tersoff-Hamann model~\cite{TH1983} confirm this
interpretation and demonstrate a good agreement of the relaxed geometries with the experimental result \cite{suppl}.

\begin{figure}[htb]
\includegraphics[width=7.5cm]{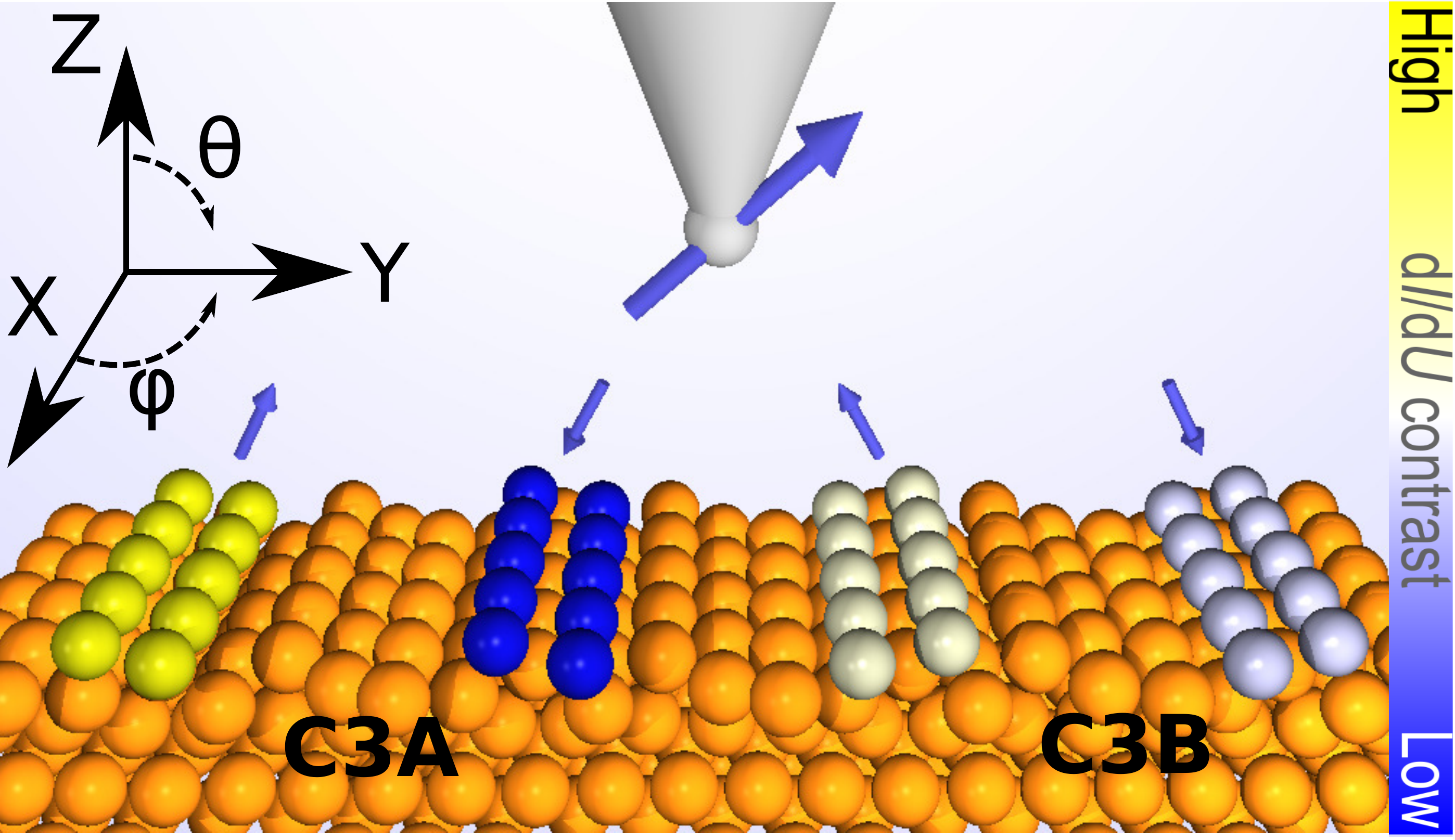}
\caption{(color online) Sketch of a magnetic tip and the two types of zigzag chains with the OH
site strand either on the right (C3A) or left (C3B).
The easy axis is canted by 33$^\circ$ from the surface normal in the direction of the OH site atom, i.e.
$\theta$=$\pm$33$^\circ$ for C3A and C3B chains, respectively. The tip is chosen at $\theta$=50$^\circ$.}
\label{fig:tipcant-v2}
\end{figure}

The anticipated SP-STM experiment is sketched in Fig.~\ref{fig:tipcant-v2}: due to the symmetry of the sample two mirror-symmetric Co zigzag chains are expected on the reconstructed Ir(001) surface, each with two possible magnetization directions along the easy axis. Within
the spin-polarized Tersoff-Hamann model \cite{Wortmann_prl_2001} the tunneling current can be written as
$I=I_0 + I_{\rm SP} {\mathbf m}_{\rm T} {\mathbf m}_{\rm S}$ where the first and second term are the non-spin-polarized and spin-polarized
contribution, respectively, and ${\mathbf m}_{\rm T}$ and ${\mathbf m}_{\rm S}$ are the unit vectors of tip and sample magnetization.
Therefore, with a perfectly
out-of-plane ($\theta = 0°$) or in-plane magnetized tip ($\theta = 90°$) only two contrast levels are measured. However, a suitable
canted tip magnetization can in principle discern all four possible magnetization directions of the chains,
as demonstrated for the tip sketched in Fig.~\ref{fig:tipcant-v2}.
As the magnetization of the tip is close to the easy axis of one type of zigzag chains (left, C3A)
there is a large positive (C3A,$\uparrow$) or negative (C3A,$\downarrow$)
contribution from the spin-polarized current for the two magnetization directions leading to a high or low d$I$/d$U$ signal,
i.e.~providing a high magnetic contrast. For the other type of zigzag chains (right, C3B) the projection of the tip magnetization
onto the chain magnetization is much smaller leading to d$I$/d$U$ signals which are closer in value.
Depending on the exact tip angle and the noise in the experiment the variation of the d$I$/d$U$ signal, i.e.~magnetization direction,
of the latter chain type will not be resolved and thus instead of a four-level contrast only a three-level contrast is obtained.

\begin{figure}[htb]
\includegraphics[width=8.5cm,angle=0]{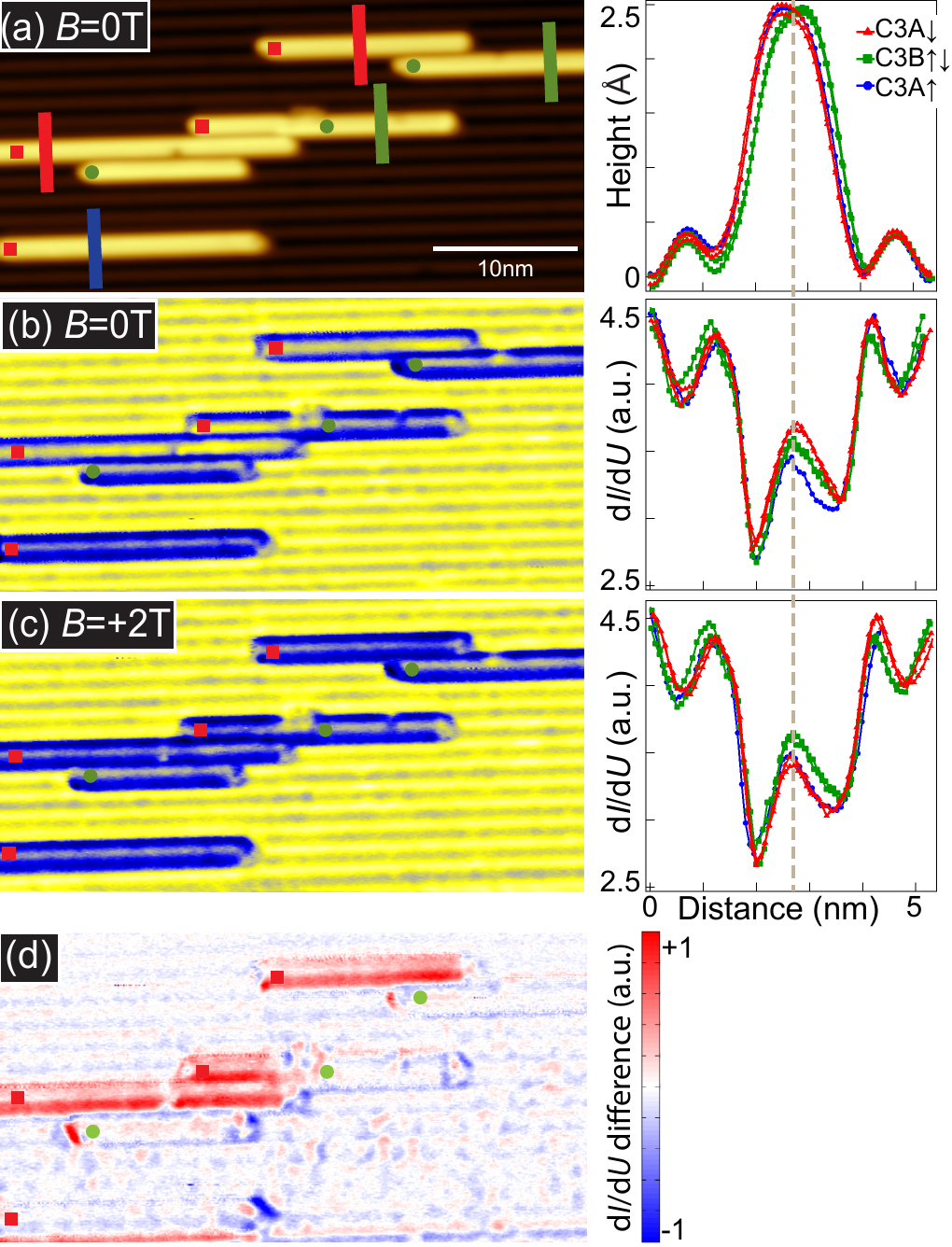}
\caption{(color online) (a) SP-STM image of zigzag Co chains which have the strand in the OH site
either on the upper (red squares, C3A) or on the lower side (green circles, C3B) of the chain (left).
Line profiles as marked in the image showing the difference between C3A and C3B structures.
(b,c) Left panels show differential conductance images taken at $B$~=~0 and +2~T respectively.
Right panels display line profiles of the d$I$/d$U$ signal taken at positions marked in (a) showing the
contrast of the C3A and C3B chains with different magnetization directions.
The dashed vertical line indicates the center of the chain. (d) difference image of the two
d$I$/d$U$ maps (b) and (c).
Measurement parameters: Cr-tip, \textit{T}~=~7.5~K, $U_{\mathrm{bias}}$~=~300~mV, $I$~=~3~nA.}
\label{fig:mag}
\end{figure}

Spin-polarized STM measurements on six zigzag Co chains are shown in Fig.~\ref{fig:mag}. The two mirror-symmetric chain configurations can be identified in the line profiles of the topography (a);
while most chains exhibit a single configuration one chain in the image area changes from C3A to C3B at a defect. The virgin state d$I$/d$U$ image in Fig.~\ref{fig:mag}(b) exhibits either a very high (yellow) or comparably low (blue) signal for the C3A-chains, while the C3B-chains show a uniform intermediate (gray) signal. This uniform contrast level on each chain is indicative of ferromagnetic order. The observation of a three level contrast
which is also visible in the line sections of Fig.~\ref{fig:mag}(b)
is in agreement with the considerations related to the sketch in Fig.~\ref{fig:tipcant-v2} confirming a canted chain
magnetization.

The application of an external magnetic field induces a magnetization reversal for the chains with antiparallel magnetization components: Fig.~\ref{fig:mag}(c) shows that the three upper C3A-chains have turned into the same magnetization state as the C3A-chain on the
lower left and they appear red in the difference image (d).
Still the magnetic contrast for the C3B-chain is in the intermediate state, leading to a two-level contrast in
applied magnetic field as seen in the right panel of Fig.~\ref{fig:mag}(c).
While the exact magnetization angle cannot be determined experimentally, we can conclude that the Co-chains are ferromagnetic with a considerable canting of the magnetization away from the high-symmetry crystallographic axes, in agreement with the theoretical findings.

In conclusion, we have demonstrated that the local breaking of symmetry of the substrate due to proximity of an atomic chain can have a gigantic effect on the direction of the chain's magnetization which is not anticipated from intuitive symmetry arguments.
Our results provide a direction for further advances in the area of control of complex low-dimensional magnets based on employing
the reduced symmetry of nano-magnets at surfaces.

We thank Matthias Menzel, Gustav Bihlmayer, and Stefan Bl\"ugel for many fruitful discussions.
J.E.B.~thanks the Alexander von Humboldt Foundation. Y.M.~gratefully acknowledges funding under the HGF-YIG programme VH-NG-513.
S.H.~thanks the Peter-Gr\"unberg Institute of the Forschungszentrum J\"ulich for its hospitality during his stay.
B.D., Y.M., F.O., and S.H.~thank the HLRN for providing computational resources and
the DFG for financial support under project HE3292/8-1.
J.E.B., A.K., K.v.B. and R.W.~thank Matthias Menzel for experimental help and
the DFG via SFB668-A8 and the ERC via the Advanced Grant FURORE for financial support.


\begin{thebibliography}{28}
\bibitem[{not({\natexlab{a}})}]{note_noco}
\eprint{Note that the ferromagnetic exchange coupling between the two Co
  strands is too strong to allow for a canting of the magnetic moments from
  each other. For a system with a weaker exchange coupling, however, such a
  non-collinear state could occur.}
\bibitem[{not({\natexlab{b}})}]{note_defect}
\eprint{The very bright chain on the middle left of the image is a
  lifted-reconstruction defect and can be ignored.}
\end{thebibliography}
\end{document}